\documentclass[conference]{IEEEtran}
\IEEEoverridecommandlockouts
\usepackage{cite}
\usepackage{amsmath,amssymb,amsfonts}
\usepackage{algorithmic}
\usepackage{graphicx}
\usepackage{textcomp}
\usepackage{booktabs}
\usepackage{multirow}
\usepackage{xcolor}
\def\BibTeX{{\rm B\kern-.05em{\sc i\kern-.025em b}\kern-.08em
    T\kern-.1667em\lower.7ex\hbox{E}\kern-.125emX}}
\begin{document}

\title{Stable Hybrid Cross-Attention Fusion for Audio-Visual Event Recognition\\
}

\author{
\IEEEauthorblockN{Parinaz Binandeh Dehaghani}
\IEEEauthorblockA{
\textit{Faculty of Engineering}\\
University of Porto\\
Porto, Portugal\\
up202100618@edu.fe.up.pt
} 
\and
\IEEEauthorblockN{Danilo Pena}
\IEEEauthorblockA{
\textit{ResoSight}\\
Montreal, Canada\\
danilo.pena@resosight.com
}
\and
\IEEEauthorblockN{A. Pedro Aguiar}
\IEEEauthorblockA{
\textit{Faculty of Engineering}\\
University of Porto\\
Porto, Portugal\\
pedro.aguiar@fe.up.pt
}

}

\maketitle

\begin{abstract}
Audio-Visual Event Recognition (AVER) is essential for intelligent urban monitoring systems, where robust multimodal understanding of complex environments is required. This paper proposes a stable hybrid cross-attention fusion framework for audio-visual event recognition in smart urban environments. The proposed architecture combines pretrained Video Masked Autoencoder (VideoMAE) and Audio Spectrogram Transformer (AST) representations with FiLM-based audio conditioning, bidirectional cross-attention fusion, multimodal Transformer encoding, and modality-temporal attention. To improve computational efficiency and training stability, frozen pretrained backbones and cached feature extraction are employed. Extensive experiments on the AVE dataset show that the proposed framework achieves the highest average performance among the evaluated unimodal and multimodal baselines across multiple evaluation metrics, obtaining a best validation accuracy of 91.74\% and a test accuracy of $83.85 \pm 1.40\%$ over five independent runs. The results indicate that the proposed hybrid fusion strategy effectively captures complementary audio-visual information and provides robust multimodal representation learning for challenging real-world urban monitoring scenarios.

\end{abstract}

\begin{IEEEkeywords}
Audio-Visual Event Recognition, Multimodal Fusion, Cross-Attention, VideoMAE, Audio Spectrogram Transformer, Smart Urban Environments
\end{IEEEkeywords}

\section{Introduction}

The rapid development of smart city technologies has significantly increased the demand for intelligent urban monitoring systems capable of understanding complex real-world environments through multimodal sensory information. Audio-Visual Event Recognition has emerged as an important research area for applications such as public safety monitoring, traffic surveillance, intelligent transportation systems, anomaly detection, and environmental awareness \cite{tian2018audio, owens2018audio, afouras2020deep}. By jointly analyzing synchronized audio and visual streams, AVER systems can provide more robust scene understanding compared with unimodal approaches.
In real-world urban environments, relying solely on visual or acoustic information is often insufficient for reliable event recognition. Visual signals may be affected by occlusions, poor illumination, motion blur, or camera viewpoint limitations, while audio signals may contain environmental noise or overlapping sound sources. Consequently, effective multimodal fusion strategies capable of exploiting complementary relationships between audio and visual modalities have become a central challenge in audio-visual learning research \cite{gao2019cooperative}.
Traditional audio-visual recognition frameworks have primarily relied on convolutional neural networks (CNNs) combined with simple feature concatenation methods for multimodal fusion. Although these approaches provide reasonable performance, they often fail to effectively model complex cross-modal dependencies and long-range temporal interactions between audio and visual streams. More recently, Transformer-based architectures have demonstrated remarkable success in computer vision and audio understanding tasks due to their ability to capture global contextual relationships through self-attention mechanisms \cite{dosovitskiy2021image, gong2021ast}.
VideoMAE has emerged as a powerful self-supervised video representation model capable of learning discriminative spatio-temporal visual features through masked video reconstruction \cite{tong2022videomae}. Similarly, the AST has demonstrated strong performance for environmental sound classification and audio event recognition using Transformer-based self-attention mechanisms \cite{gong2021ast}. Despite these advances, effectively integrating pretrained audio and visual Transformer representations while maintaining computational efficiency and stable optimization remains a challenging problem, particularly for multimodal learning in smart urban environments. 

To address these limitations, this paper proposes a stable hybrid cross-attention fusion framework for audio-visual event recognition in smart urban environments. The proposed framework combines pretrained VideoMAE and AST representations with FiLM-based audio conditioning, bidirectional cross-attention fusion, multimodal Transformer encoding, and modality-temporal attention. In addition, frozen pretrained backbones and cached feature extraction are employed to improve computational efficiency and training stability. 
The main contributions of this work are as follows:
\begin{itemize}
\item A cached-feature audio-visual recognition framework combining frozen VideoMAE and AST embeddings for efficient training on medium-scale AVE (Audio-Visual Event) datasets. 
\item A hybrid fusion module combining FiLM-based audio conditioning, stable bidirectional cross-attention refinement, multimodal Transformer encoding, and modality-temporal attention for enhanced cross-modal interaction.
\item A multi-seed empirical evaluation on the AVE dataset against audio-only, video-only, and simple multimodal fusion baselines, including accuracy, balanced accuracy, macro/weighted F1, and computational cost.
\end{itemize}
This paper is organized as follows. Section II reviews related work in audio-visual event recognition and multimodal fusion. Section III presents the proposed methodology, including feature extraction, multimodal fusion, and training strategies. Section IV describes the experimental setup, dataset, implementation details, and evaluation protocol. Section V discusses the experimental results and computational analysis. Finally, Section VI concludes the paper and outlines future research directions.

\begin{figure*}[!t]
\centering
\includegraphics[width=\textwidth]{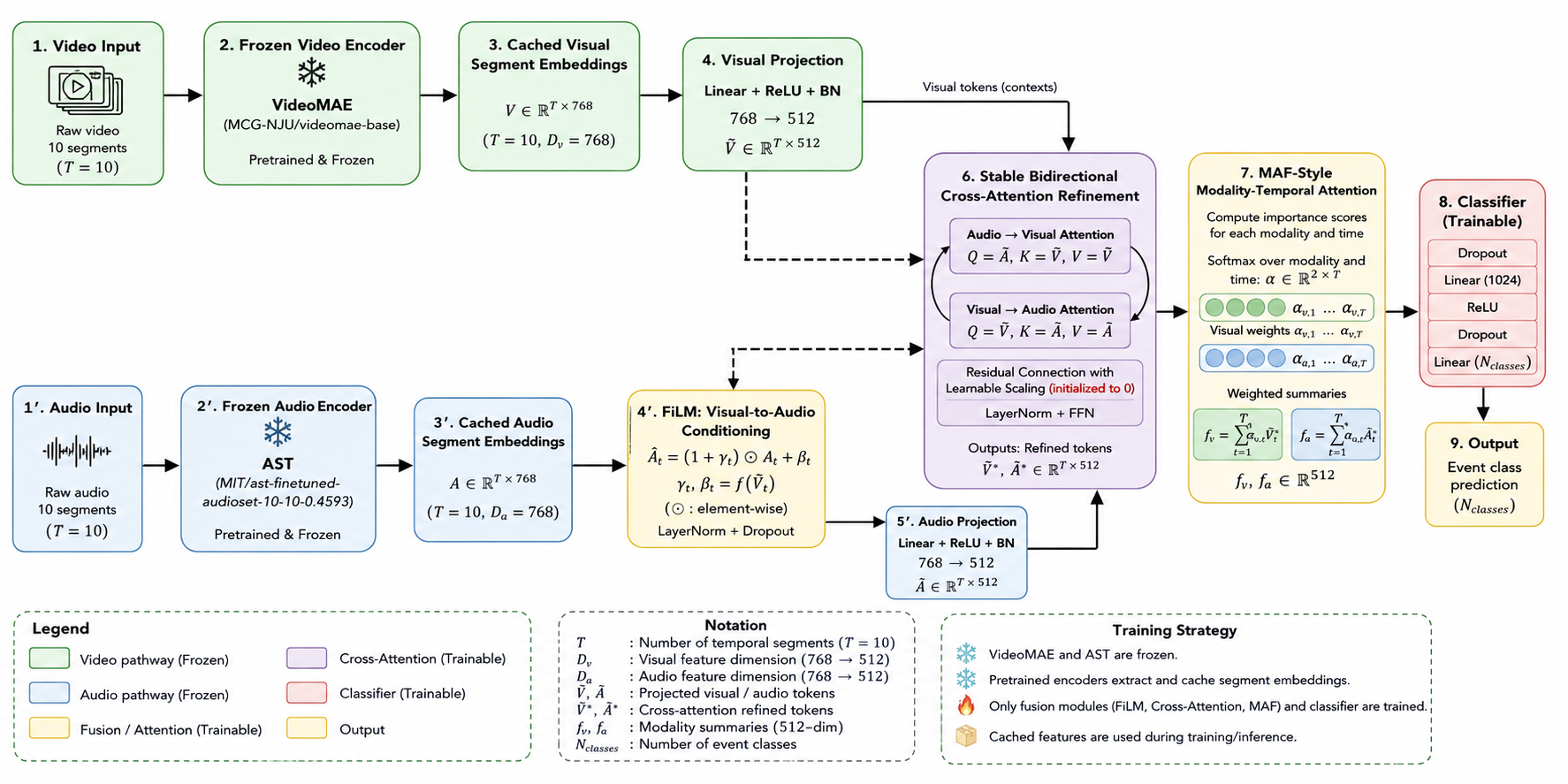}
\caption{Overview of the proposed Hybrid Fusion framework for Audio-Visual Event Recognition (AVER).}
\label{fig:model}
\end{figure*}
\section{RELATED WORK}

Audio-visual event recognition has become an important research topic in multimodal scene understanding, particularly for intelligent surveillance, smart urban monitoring, anomaly detection, and environmental awareness. Compared with unimodal approaches, audio-visual learning can exploit complementary information from synchronized acoustic and visual streams, leading to more robust recognition in complex real-world environments.
Tian \textit{et al.}~\cite{tian2018audio} introduced the AVE dataset and formulated audio-visual event localization in unconstrained videos. Their work defined an audio-visual event as an event that is both visible and audible in a video segment and demonstrated the importance of jointly modeling audio and visual modalities. They further showed that temporal alignment and multimodal fusion are essential for effective audio-visual event localization.
Several studies have investigated different audio-visual fusion strategies for event recognition. Brousmiche \textit{et al.}~\cite{brousmiche2019fusion} explored audio-visual fusion and conditioning with neural networks and showed that multimodal event classification generally outperforms unimodal classification. Their study investigated different fusion and conditioning mechanisms, highlighting the importance of modality interaction in audio-visual event recognition.

Building on this direction, Brousmiche \textit{et al.}~\cite{brousmiche2022mafnet} proposed the Multimodal Attentive Fusion Network (MAFnet), which dynamically fuses audio and visual information across temporal segments. MAFnet introduced a modality-temporal attention mechanism to assign different importance scores to audio and visual features over time. This approach demonstrated that audio-visual event recognition benefits from dynamically selecting the most informative modality and temporal window.
Audio-visual representation learning has also been explored in safety-critical monitoring scenarios. Gao \textit{et al.}~\cite{gao2024anomaly} investigated audio-visual representation learning for anomaly event detection in crowds. Their work emphasized that audio signals can provide useful complementary information when visual cues are limited by occlusion or crowd density. This observation is particularly relevant to smart urban environments, where robust event recognition often requires multimodal perception.
Recent advances in Transformer-based models have further improved representation learning for both video and audio understanding. VideoMAE~\cite{tong2022videomae} introduced a masked autoencoding strategy for self-supervised video pretraining and demonstrated strong spatio-temporal representation learning. In the audio domain, the Audio Spectrogram Transformer ~\cite{gong2021ast} applied Transformer-based self-attention to spectrogram representations and achieved strong performance in audio classification tasks.

Despite these advances, many existing audio-visual event recognition methods rely on relatively shallow fusion strategies, such as concatenation, score-level fusion, or modality weighting. While effective, these methods may not fully capture complex cross-modal dependencies between audio and visual temporal representations. Moreover, end-to-end training of large audio-visual Transformer models can be computationally expensive and unstable, particularly for medium-scale datasets such as AVE.
To address these limitations, the proposed framework combines frozen pretrained VideoMAE and AST representations with FiLM-based audio conditioning, bidirectional cross-attention fusion, multimodal Transformer encoding, and modality-temporal attention. This design extends previous audio-visual fusion approaches by explicitly modeling cross-modal dependencies while maintaining computational efficiency through cached feature extraction.

\section{METHODOLOGY} 

The proposed framework aims to perform robust audio-visual event recognition through stable multimodal representation learning and hybrid cross-attention fusion. The overall architecture consists of four major stages: visual feature extraction, audio feature extraction, FiLM-based audio conditioning, and stable hybrid cross-attention fusion with modality-temporal attention. An overview of the complete architecture is illustrated in Fig.~\ref{fig:model}, which provides a schematic representation of the data flow and interactions between the different modules.
The following subsections describe each component of the framework in detail, including the feature extraction pipelines, cross-modal conditioning mechanism, hybrid fusion architecture, and training strategy.

\subsection{Video Feature Extraction}

To obtain robust spatio-temporal visual representations for audio-visual event recognition, the proposed framework employs a pretrained VideoMAE as the visual backbone. VideoMAE is a transformer-based self-supervised model capable of learning discriminative temporal and spatial visual patterns through masked video reconstruction.
Each input video is divided into $T=10$ synchronized temporal segments, and a fixed number of frames are uniformly sampled from each segment. The sampled frames are resized, normalized, and processed by the pretrained VideoMAE encoder to generate segment-level visual embeddings.
The final segment representation is obtained by averaging the latent transformer token embeddings.
Formally, let a video be divided into $T=10$ synchronized temporal segments, and let $X_t^{(v)}$ denote the set of sampled frames from the $t$-th segment, for $t \in \{1,\dots,T\}$. Each segment is processed by the pretrained VideoMAE encoder, which maps the input frames into a sequence of latent token embeddings:
\begin{equation}
Z_t^{(v)} =
\mathrm{VideoMAE}\!\left(X_t^{(v)}\right)
\in \mathbb{R}^{N_v \times d_v},
\end{equation}
where $N_v$ denotes the number of output tokens and $d_v$ is the latent embedding dimension.
To obtain a compact segment-level visual representation, the latent token embeddings are aggregated by average pooling:
\begin{equation}
v_t =
\frac{1}{N_v}
\sum_{i=1}^{N_v}
Z_{t,i}^{(v)}
\in \mathbb{R}^{d_v},
\end{equation}
where $Z_{t,i}^{(v)}$ denotes the $i$-th token embedding of the $t$-th temporal segment.
The final visual representation of the input video is therefore given by the ordered sequence of segment-level embeddings:
\begin{equation}
V =
\{v_1, v_2, \dots, v_T\}
\in \mathbb{R}^{T \times d_v},
\end{equation}
which preserves the temporal structure of the video while providing discriminative spatio-temporal features for the subsequent multimodal fusion stage.

\subsection{Audio Feature Extraction}

To obtain robust acoustic representations, the proposed framework employs the AST as the audio feature extraction backbone. AST is a transformer-based architecture capable of modeling long-range temporal dependencies through self-attention mechanisms.
For each temporal segment, the corresponding audio waveform is extracted, resampled to 16 kHz, and transformed into a log-Mel spectrogram representation. The spectrogram is then processed by the pretrained AST encoder to generate segment-level audio embeddings. The final audio representation is obtained by averaging the latent transformer token embeddings.

\subsection{FiLM-Based Audio Conditioning}

To improve cross-modal interaction prior to multimodal fusion, the proposed framework employs a Feature-wise Linear Modulation (FiLM) mechanism that conditions the audio representations on the corresponding visual embeddings. Given the segment-level visual and audio representations $v_t \in \mathbb{R}^{d}$ and $a_t \in \mathbb{R}^{d}$, the FiLM module generates adaptive, feature-wise affine parameters from the visual modality:
\begin{equation}
\gamma_t = W_{\gamma} v_t + b_{\gamma},
\qquad
\beta_t = W_{\beta} v_t + b_{\beta},
\end{equation}
where $W_{\gamma}, W_{\beta} \in \mathbb{R}^{d \times d}$ and $b_{\gamma}, b_{\beta} \in \mathbb{R}^{d}$ are learnable parameters.
The conditioned audio representation $\hat{a}_t \in \mathbb{R}^{d}$ is then obtained by applying feature-wise affine modulation:
\begin{equation}
\hat{a}_t =
(1 + \gamma_t) \odot a_t + \beta_t,
\end{equation}
where $\odot$ denotes element-wise multiplication.
This formulation allows each feature dimension of the audio embedding to be adaptively scaled and shifted according to the visual context, introducing both multiplicative and additive cross-modal interactions. As a result, the model can emphasize acoustically relevant patterns that are consistent with the visual input while suppressing irrelevant or noisy components. Compared with simple concatenation-based fusion, this conditioning mechanism provides a structured and parameter-efficient way to inject cross-modal information prior to the attention-based fusion stage, thereby improving the discriminative power of the learned multimodal representations.

\subsection{Stable Hybrid Cross-Attention Fusion}

After FiLM-based conditioning, the proposed framework performs multimodal fusion through a hybrid cross-attention mechanism designed to capture bidirectional interactions between audio and visual representations. Let $V=\{v_1,\dots,v_T\}\in\mathbb{R}^{T\times d}$ and $\hat{A}=\{\hat{a}_1,\dots,\hat{a}_T\}\in\mathbb{R}^{T\times d}$ denote the visual and conditioned audio sequences, respectively.
The scaled dot-product attention mechanism is defined as:
\begin{equation}
\mathrm{Attn}(Q,K,V)
=
\mathrm{softmax}
\left(
\frac{QK^{\top}}{\sqrt{d}}
\right)V,
\end{equation}
where $Q$, $K$, and $V$ denote the query, key, and value matrices, and $d$ is the embedding dimension.
\textit{Audio-to-visual attention.}
The conditioned audio sequence attends to the visual sequence:
\begin{equation}
Q_a = \hat{A}W_Q^{(a)}, \quad
K_v = VW_K^{(v)}, \quad
V_v = VW_V^{(v)},
\end{equation}
\begin{equation}
H_a = \mathrm{Attn}(Q_a,K_v,V_v),
\end{equation}
where $W_Q^{(a)}$, $W_K^{(v)}$, and $W_V^{(v)}$ are learnable projection matrices, and $H_a\in\mathbb{R}^{T\times d}$ denotes the refined audio representation.
\textit{Visual-to-audio attention.}
Similarly, the visual sequence attends to the conditioned audio sequence:
\begin{equation}
Q_v = VW_Q^{(v)}, \quad
K_a = \hat{A}W_K^{(a)}, \quad
V_a = \hat{A}W_V^{(a)},
\end{equation}
\begin{equation}
H_v = \mathrm{Attn}(Q_v,K_a,V_a),
\end{equation}
where $W_Q^{(v)}$, $W_K^{(a)}$, and $W_V^{(a)}$ are learnable projection matrices, and $H_v\in\mathbb{R}^{T\times d}$ denotes the refined visual representation.
To improve training stability and representation quality, residual connections, layer normalization, and feed-forward refinement layers are applied after the bidirectional cross-attention stage. The refined audio and visual representations are subsequently concatenated and processed by a multimodal Transformer encoder, which jointly models temporal dependencies and cross-modal correlations through self-attention and feed-forward transformations. This additional encoding stage enables more effective integration of complementary audio-visual information and improves the robustness of the learned multimodal representations.
 
\subsection{Modality-Temporal Attention}

After cross-attention fusion and Transformer encoding, a modality-temporal attention mechanism is applied to aggregate the most informative audio-visual tokens. Let $H''=\{h_1,\dots,h_{2T}\}$ denote the encoded multimodal token sequence. The attention weights are computed as:
\begin{equation}
\alpha_i =
\frac{\exp(e_i)}
{\sum_{j=1}^{2T}\exp(e_j)},
\qquad
e_i = w^\top \tanh(Wh_i+b).
\end{equation}
The final fused representation is obtained by weighted aggregation:
\begin{equation}
f =
\sum_{i=1}^{2T}
\alpha_i h_i.
\end{equation}
This mechanism enables the model to adaptively emphasize informative modalities and temporal segments while suppressing noisy or irrelevant tokens.

\subsection{Stable Training Strategy}

To improve computational efficiency, the pretrained VideoMAE and AST backbones are kept frozen during training, while only the fusion and classification modules are optimized. Furthermore, segment-level audio and visual embeddings are extracted offline and stored in a cached feature repository, allowing the model to operate directly on precomputed representations.
Given the final fused representation $f$, the predicted class probabilities are obtained through a softmax classifier:
\begin{equation}
\hat{y}=
\mathrm{softmax}(W_c f+b_c).
\end{equation}
To address class imbalance, a class-balanced cross-entropy loss is employed:
\begin{equation}
\mathcal{L}
=
-
\sum_{c=1}^{C}
w_c y_c \log(\hat{y}_c),
\end{equation}
where $w_c$ denotes the class weight associated with class $c$.
Training stability is further enhanced using automatic mixed precision, dropout regularization, residual connections, layer normalization, adaptive learning-rate scheduling, and early stopping. All experiments are conducted using five independent random seeds, and the final results are reported as mean $\pm$ standard deviation.

\section{Experimental Setup}

\subsection{Dataset}

The proposed framework is evaluated on the AVE dataset, which is a widely used benchmark for audio-visual event recognition tasks. The AVE dataset contains temporally aligned audio and visual event annotations collected from unconstrained real-world videos spanning multiple event categories.
Each video clip has a duration of approximately 10 seconds and is associated with synchronized audio and visual streams. Following the official dataset protocol, the videos are divided into training, validation, and test subsets. During preprocessing, corrupted or missing video files are automatically discarded to ensure stable training and evaluation.
To preserve temporal consistency, each video is divided into $T=10$ non-overlapping temporal segments, where each segment corresponds to one second of audio-visual content.

\subsection{Implementation Details}

The proposed framework is implemented in PyTorch and trained on an NVIDIA T4 GPU. Pretrained VideoMAE (\texttt{MCG-NJU/videomae-base}) and AST (\texttt{MIT/ast-finetuned-audioset-10-10-0.4593}) models are used as frozen visual and audio feature extractors. To improve computational efficiency, audio and visual embeddings are extracted offline and stored in a cached feature repository, such that only the fusion and classification modules are optimized during training.
The fusion module employs a hidden dimension of 512, 8 attention heads, and 2 Transformer layers. Dropout with a probability of 0.5 is applied in the classifier. The model is trained using AdamW with an initial learning rate of $10^{-3}$ and weight decay of $10^{-4}$. A ReduceLROnPlateau scheduler, mixed-precision training, and early stopping are employed to improve optimization efficiency and generalization. Training is performed with a batch size of 4 for up to 80 epochs.

\subsection{Evaluation Protocol and Baseline Models}
The proposed framework is evaluated using multiple performance metrics, including accuracy, balanced accuracy, precision, recall, and F1-score. Both macro-averaged and weighted-averaged variants of precision, recall, and F1-score are reported to account for class imbalance and class frequency distributions. To ensure robustness, all experiments are conducted using five independent random seeds, and results are reported as mean $\pm$ standard deviation.
Four architectures are evaluated. The \textbf{Video-only} baseline utilizes pretrained VideoMAE visual embeddings without audio information, while the \textbf{Audio-only} baseline employs pretrained AST audio embeddings only. The \textbf{Simple AV Fusion} baseline combines FiLM-based audio conditioning with feature concatenation. Finally, the \textbf{Hybrid Fusion} framework incorporates FiLM conditioning, bidirectional cross-attention, multimodal Transformer encoding, and modality-temporal attention. This comparison enables analysis of the contribution of each modality and fusion strategy to audio-visual event recognition performance.

\section{Results and Discussion} 

\begin{table*}[t]
\centering
\caption{Multi-seed performance comparison over five random seeds.}
\label{tab:multiseed_results}
\resizebox{\textwidth}{!}{%
\begin{tabular}{lcccc}
\toprule
\textbf{Metric} &
\textbf{Proposed Hybrid Fusion} &
\textbf{Simple AV Fusion} &
\textbf{Audio-only AST} &
\textbf{Video-only VideoMAE} \\
\midrule
best\_val\_acc      & $0.8948 \pm 0.0084$ & $0.8843 \pm 0.0146$ & $0.8670 \pm 0.0085$ & $0.6035 \pm 0.0135$ \\
test\_loss          & $0.8853 \pm 0.0274$ & $0.8923 \pm 0.0887$ & $1.0489 \pm 0.0849$ & $1.9214 \pm 0.0785$ \\
test\_acc           & $0.8385 \pm 0.0140$ & $0.8239 \pm 0.0140$ & $0.7991 \pm 0.0100$ & $0.5060 \pm 0.0215$ \\
balanced\_accuracy  & $0.8277 \pm 0.0168$ & $0.8228 \pm 0.0231$ & $0.7725 \pm 0.0127$ & $0.5038 \pm 0.0304$ \\
macro\_f1           & $0.8210 \pm 0.0196$ & $0.8124 \pm 0.0274$ & $0.7579 \pm 0.0127$ & $0.4909 \pm 0.0312$ \\
weighted\_f1        & $0.8368 \pm 0.0145$ & $0.8215 \pm 0.0152$ & $0.7935 \pm 0.0092$ & $0.5024 \pm 0.0229$ \\
macro\_precision    & $0.8447 \pm 0.0144$ & $0.8421 \pm 0.0222$ & $0.7822 \pm 0.0171$ & $0.5252 \pm 0.0360$ \\
weighted\_precision & $0.8587 \pm 0.0085$ & $0.8483 \pm 0.0085$ & $0.8210 \pm 0.0127$ & $0.5498 \pm 0.0271$ \\
macro\_recall       & $0.8277 \pm 0.0168$ & $0.8228 \pm 0.0231$ & $0.7725 \pm 0.0127$ & $0.5038 \pm 0.0304$ \\
weighted\_recall    & $0.8385 \pm 0.0140$ & $0.8239 \pm 0.0140$ & $0.7991 \pm 0.0100$ & $0.5060 \pm 0.0215$ \\
\bottomrule
\end{tabular}
}
\end{table*}

\begin{figure}[t]
    \centering
    \includegraphics[width=0.48\textwidth]{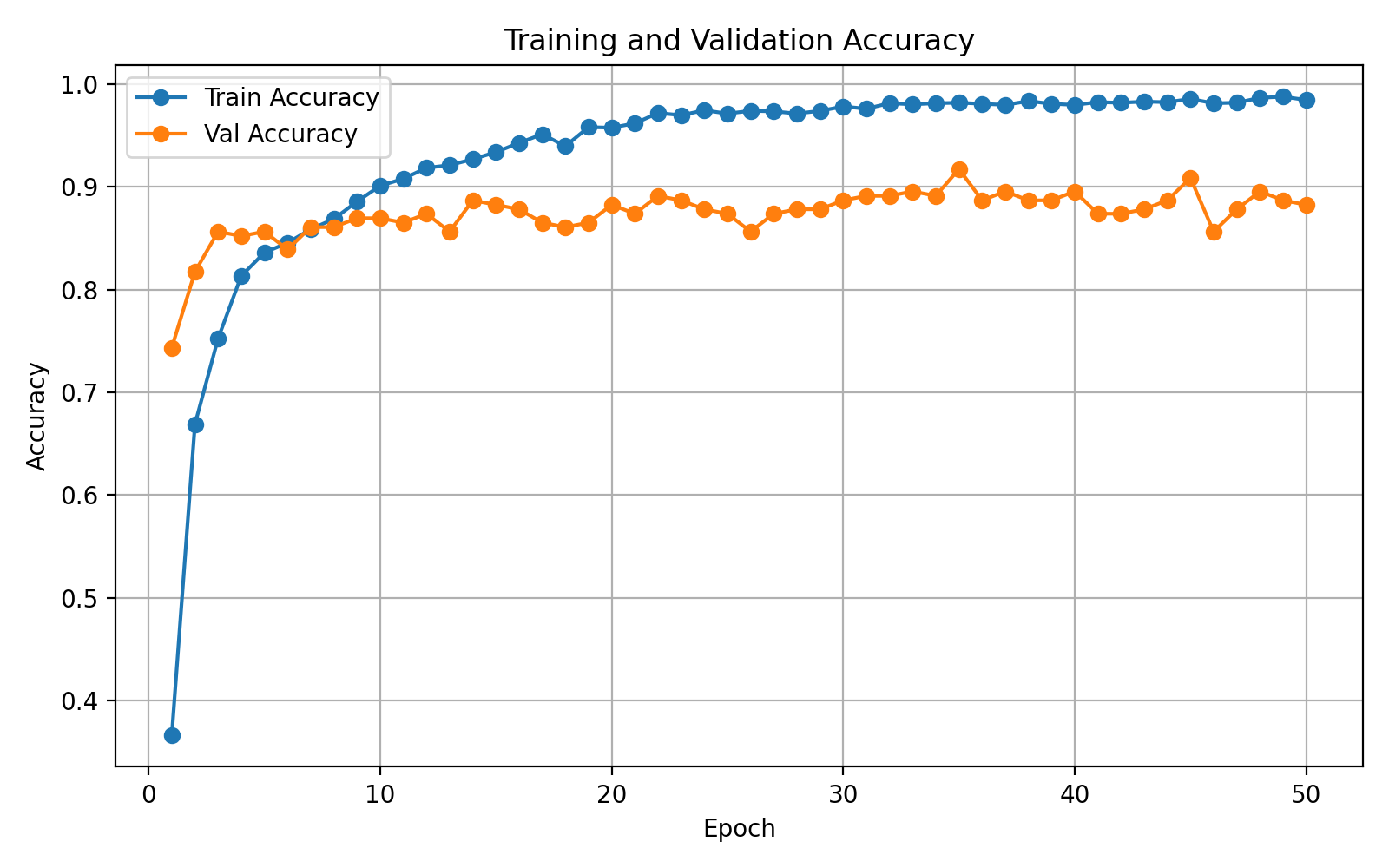}
    \caption{Training and validation accuracy of the proposed Hybrid Fusion model.}
    \label{fig:accuracy_curve}
\end{figure}
\begin{figure}[t]
    \centering
    \includegraphics[width=0.48\textwidth]{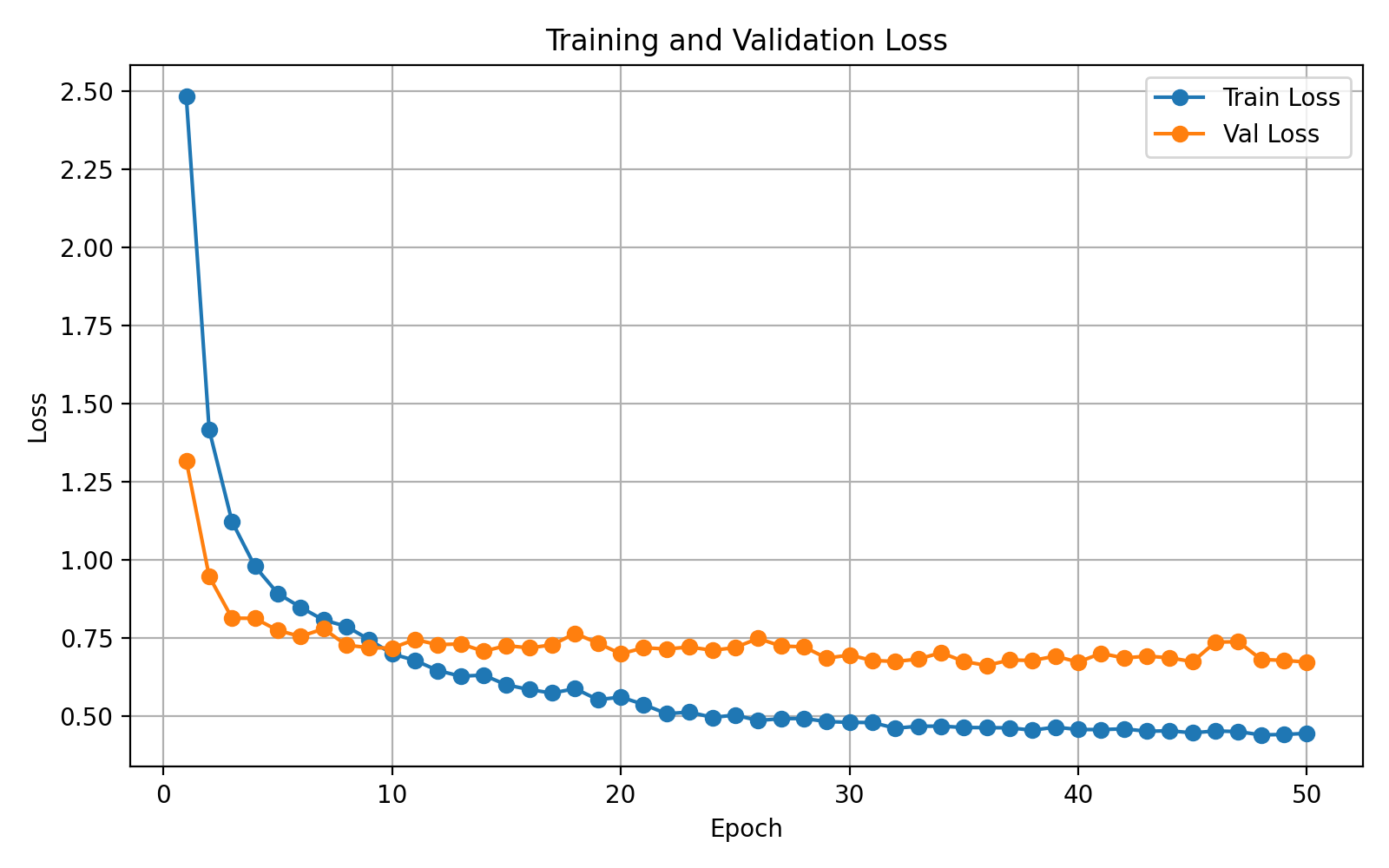}
    \caption{Training and validation loss of the proposed Hybrid Fusion model.}
    \label{fig:loss_curve}
\end{figure}
\begin{figure}[t]
    \centering
    \includegraphics[width=0.48\textwidth]{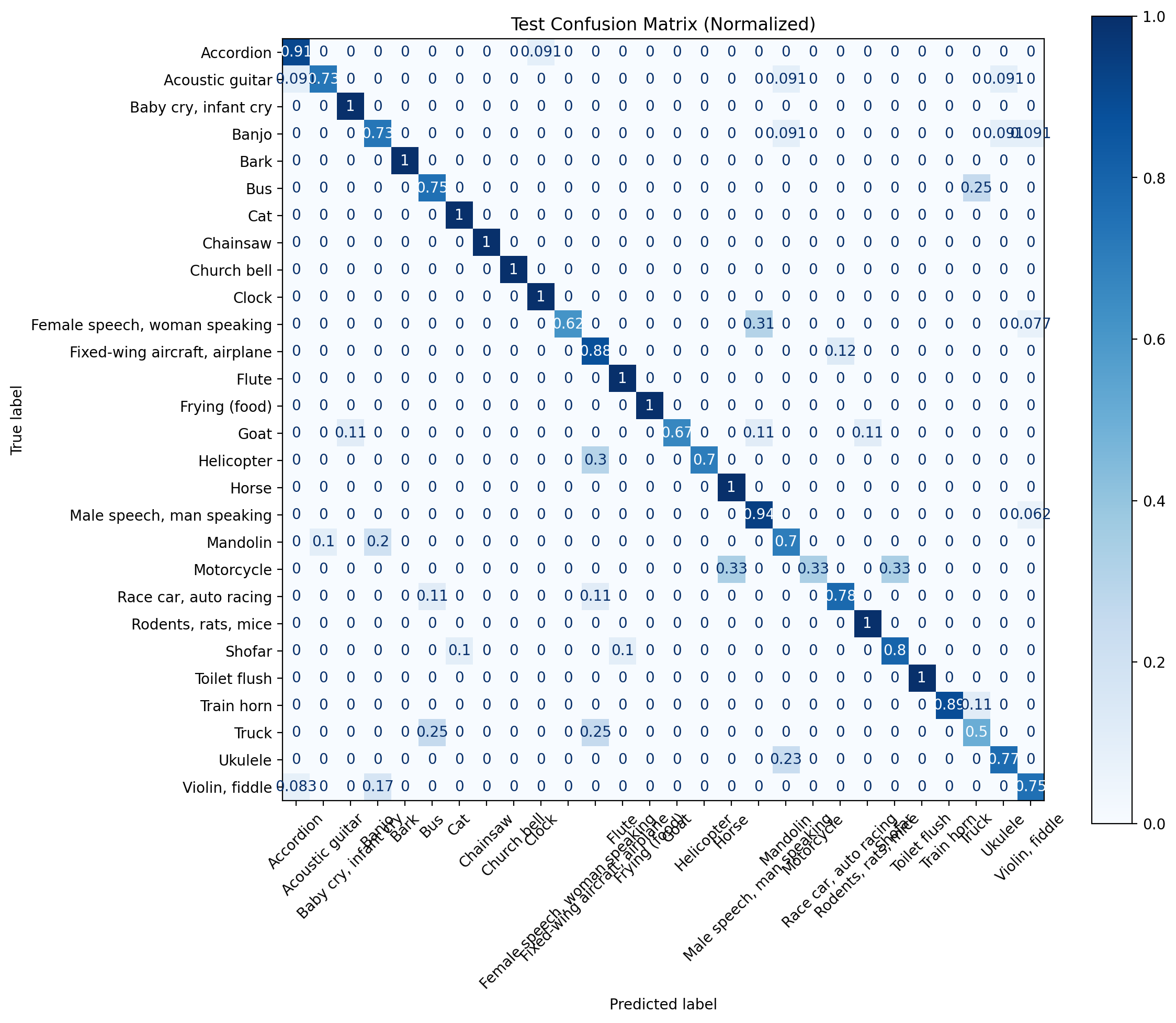}
    \caption{Normalized confusion matrix of the proposed Hybrid Fusion model on the AVE test set.}
    \label{fig:confusion_matrix}
\end{figure}

\begin{table*}[t]
\centering
\caption{Computational cost comparison of the evaluated models.}
\label{tab:computational_cost}
\resizebox{\textwidth}{!}{
\begin{tabular}{lccccc}
\hline
\textbf{Model} & \textbf{Fusion Type} & \textbf{Params (M)} & \textbf{Training Time (h)} & \textbf{GPU} & \textbf{Usable Split Sizes After Cache Filtering} \\
\hline
Audio-only AST 
& Audio-only temporal attention 
& 0.804381 
& 0.5175 
& T4 GPU 
& train=1891 $\mid$ val=230 $\mid$ test=234 \\

Video-only VideoMAE 
& Video-only 
& 0.804381 
& 0.2923 
& T4 GPU 
& train=1891 $\mid$ val=230 $\mid$ test=234 \\

Simple AV Fusion 
& Concatenation 
& 2.643 
& 0.3134 
& T4 GPU 
& train=1891 $\mid$ val=230 $\mid$ test=234 \\

Proposed Hybrid Fusion 
& FiLM + Cross-Attention + MAF Attention 
& 6.856 
& 0.3835 
& T4 GPU 
& train=1891 $\mid$ val=230 $\mid$ test=234 \\
\hline
\end{tabular}
}
\end{table*} 

This section presents the performance evaluation of the proposed framework on the AVE dataset. Table~\ref{tab:multiseed_results} reports the mean $\pm$ standard deviation over five independent runs. The results show that multimodal fusion generally achieves higher average performance than unimodal approaches. While the Video-only VideoMAE baseline yields the weakest results, the Audio-only AST baseline demonstrates the strong discriminative capability of acoustic representations. The Simple AV Fusion baseline further improves performance through multimodal integration.
Among the evaluated methods, the proposed Hybrid Fusion framework achieves the highest average performance across most metrics, suggesting that bidirectional cross-attention, multimodal Transformer encoding, and modality-temporal attention facilitate more effective multimodal representation learning.

Figures~\ref{fig:accuracy_curve} and \ref{fig:loss_curve} illustrate stable training behavior, while the confusion matrix in Fig.~\ref{fig:confusion_matrix} indicates strong classification performance across most event categories, with some expected confusion between acoustically or visually similar classes.
Table~\ref{tab:computational_cost} summarizes the computational cost of the evaluated models. Although the proposed Hybrid Fusion framework contains the largest number of trainable parameters, its training time remains practical due to the use of frozen pretrained backbones and cached feature extraction. Overall, the results suggest that the proposed framework provides a favorable balance between recognition performance and computational efficiency.

\section{Conclusion}
The experimental results demonstrated that the proposed \textbf{Hybrid Fusion} framework consistently achieved the best overall performance across nearly all evaluation metrics compared with both unimodal and conventional multimodal baselines. In particular, the proposed framework achieved $0.8385 \pm 0.0140$ test accuracy, $0.8277 \pm 0.0168$ balanced accuracy, and $0.8368 \pm 0.0145$ weighted F1-score, outperforming the Simple AV Fusion, Audio-only AST, and Video-only VideoMAE baselines.
The experimental analysis additionally revealed that the \textbf{Video-only VideoMAE} baseline achieved the lowest overall performance despite having the shortest training time, indicating that visual information alone is insufficient for robust recognition of many AVE categories. In contrast, the \textbf{Audio-only AST} baseline provided substantially stronger performance, demonstrating the highly discriminative capability of acoustic representations for environmental event recognition tasks.
The \textbf{Simple AV Fusion} baseline further improved the performance by combining audio and visual modalities through FiLM-based conditioning and feature concatenation. However, the proposed \textbf{Hybrid Fusion} framework consistently achieved superior results owing to the incorporation of bidirectional cross-attention, multimodal Transformer fusion, and modality-temporal attention mechanisms. Unlike simple concatenation-based fusion, the proposed architecture explicitly models cross-modal dependencies and dynamically learns complementary relationships between audio and visual temporal representations.

Furthermore, the relatively low standard deviation across multiple random seeds demonstrates that the proposed framework not only improves recognition performance but also provides robust and stable multimodal representation learning. Although the proposed architecture introduces additional model complexity due to the cross-attention and multimodal Transformer modules, the increase in training time remains relatively small compared with the Simple AV Fusion baseline, increasing only from $0.3134$ hours to $0.3835$ hours on a single NVIDIA T4 GPU. Therefore, the additional computational complexity is well justified by the achieved gains in audio-visual event recognition performance.


\bibliographystyle{IEEEtran}
\bibliography{references}

@inproceedings{tian2018audio,
  title={Audio-Visual Event Localization in Unconstrained Videos},
  author={Tian, Yapeng and Shi, Jing and Li, Bochen and Duan, Zhiyao and Xu, Chenliang},
  booktitle={Proceedings of the European Conference on Computer Vision (ECCV)},
  year={2018}
}

@inproceedings{owens2018audio,
  title={Audio-Visual Scene Analysis with Self-Supervised Multisensory Features},
  author={Owens, Andrew and Efros, Alexei A.},
  booktitle={Proceedings of the European Conference on Computer Vision (ECCV)},
  year={2018}
}

@article{afouras2020deep,
  title={Deep Audio-Visual Speech Recognition},
  author={Afouras, Triantafyllos and Chung, Joon Son and Senior, Andrew and Vinyals, Oriol and Zisserman, Andrew},
  journal={IEEE Transactions on Pattern Analysis and Machine Intelligence},
  volume={44},
  number={12},
  pages={8717--8727},
  year={2020},
  publisher={IEEE}
}

@inproceedings{gao2019cooperative,
  title={Cooperative Learning of Audio and Video Models from Self-Supervised Synchronization},
  author={Gao, Ruohan and Grauman, Kristen},
  booktitle={Advances in Neural Information Processing Systems (NeurIPS)},
  volume={32},
  year={2019}
}

@inproceedings{dosovitskiy2021image,
  title={An Image is Worth 16x16 Words: Transformers for Image Recognition at Scale},
  author={Dosovitskiy, Alexey and Beyer, Lucas and Kolesnikov, Alexander and others},
  booktitle={International Conference on Learning Representations (ICLR)},
  year={2021}
}

@inproceedings{gong2021ast,
  title={AST: Audio Spectrogram Transformer},
  author={Gong, Yuan and Chung, Yu-An and Glass, James},
  booktitle={Proceedings of Interspeech},
  year={2021}
}

@inproceedings{tong2022videomae,
  title={VideoMAE: Masked Autoencoders are Data-Efficient Learners for Self-Supervised Video Pre-Training},
  author={Tong, Zhan and Song, Yibing and Wang, Jue and Wang, Limin},
  booktitle={Advances in Neural Information Processing Systems (NeurIPS)},
  volume={35},
  pages={10078--10093},
  year={2022}
}

@inproceedings{brousmiche2019fusion,
  title={Audio-Visual Fusion and Conditioning with Neural Networks for Event Recognition},
  author={Brousmiche, Mathilde and Rouat, Jean and Dupont, St{\'e}phane},
  booktitle={2019 IEEE 29th International Workshop on Machine Learning for Signal Processing (MLSP)},
  pages={1--6},
  year={2019},
  organization={IEEE}
}

@article{brousmiche2022mafnet,
  title={Multimodal Attentive Fusion Network for Audio-Visual Event Recognition},
  author={Brousmiche, Mathilde and Rouat, Jean and Dupont, St{\'e}phane},
  journal={Information Fusion},
  volume={85},
  pages={52--59},
  year={2022},
  publisher={Elsevier}
}

@article{gao2024anomaly,
  title={Audio--Visual Representation Learning for Anomaly Events Detection in Crowds},
  author={Gao, Junyu and Gong, Maoguo and Li, Xuelong},
  journal={Neurocomputing},
  volume={582},
  pages={127489},
  year={2024},
  publisher={Elsevier}
}

\end{document}